# THz beam steering by optical coherent control

Heiko Füser[1] and Mark Bieler[1]

[1]Physikalisch-Technische Bundesanstalt, Bundesallee 100, 38116 Braunschweig, Germany

We demonstrate optical coherent control of the emission direction of THz radiation. Femtosecond laser pulses are used to excite different types of ultrafast photocurrents along different directions in a bulk GaAs sample. The overall emission pattern can be modified by changing the phase of the optical excitation. With this method, THz beam steering of about 8 degrees is realized. A simple dipole-based model allows us to relate the size of the steering effect to the amplitude ratio between the different photocurrent contributions and to diffraction effects resulting from the excitation spot size.

Research and development in the THz frequency range has been significantly intensified in many scientific and technical areas during the last decades. The main drivers for this process were existing and envisaged applications in communications, spectroscopy, material science, and security[1]. Among these drivers, the need for spatial control has become more and more important. For instance, communications at high frequencies requires the realization of different propagation channels[2]. Furthermore, spatially-resolved spectroscopy and imaging applications are based on a relative movement between the THz field and the device under test (DUT)[3]. One possible approach to accomplish such demands is given by a steerable THz emitter. Various technical realizations based on changes of laser excitation angles or diffractive gratings have been introduced so far[4–8]. However, mechanical movement of components within a measurement setup limits the speed of the measurement process and diffraction effects usually show strong frequency-dependant steering angles. Thus, the development of techniques that allow for non mechanical and broadband steering of the THz beam is desirable.

In this work, we present an alternative approach for fast and reliable control of the emission direction of THz radiation. The beam steering as demonstrated here does not rely on mechanical movements of any parts of the setup or on beam shaping apertures but is realized by optical coherent control, i.e., by modifying the phase of the optical excitation pulse. Beam steering is realized by controlling the interference between different optically-induced currents in a bulk GaAs sample. Due to the nonlinear properties of the GaAs crystal, it is possible to induce shift currents, which result from a spatial shift of the center of the electron charge during excitation[9]. Those currents depend on the phase of the excitation pulse and, along certain crystal axes, the direction can be reversed by changing the phase of the excitation. Additionally, photoexcited free carriers are accelerated perpendicular to the crystal's surface by the surface depletion field[10], which results from band-structure bending at the crystal/air interface. We experimentally and theoretically show below that the overall THz radiation emitted from both current contributions forms an emission pattern whose direction can be steered by coherently controlling the shift current direction. In our proof-of-concept experiments, steering angles up to 8 degrees are realized, yet, simulations show that the steering angle can also be significantly enhanced.

The experiments were performed on semi-insulating bulk GaAs samples with (110) and (113) crystal orientations, which were excited by normally-incident and linearly-polarized optical pulses derived from a Ti:sapphire laser (repetition rate 76 MHz, pulse length 100 fs, center wavelength 820 nm). The generation of shift currents can be described using a third-rank shift current tensor $\overleftrightarrow{\sigma}$, reflecting the crystal's point group symmetry and the polarization direction of the excitation. The shift current tensor is purely real and symmetric in the last two Cartesian indices, i.e., $\sigma_{abc} = \sigma_{acb}$[9]. In this notation, the current flow direction is described by the first index, while the second and third indices indicate the direction of the applied electrical field vectors. In the following we take the [001], [1$\bar{1}$0], and [110] axes of the (110)-oriented GaAs sample as the $x$, $y$, and $z$ axes, respectively. Focusing on the $\sigma_{yxy} = \sigma_{yyx}$ shift current tensor elements, the current flow along $y$ is described by[9]

$$J_y^{Shift} = 4\, \sigma_{yxy}\, E_x E_y \cos(\varphi_x - \varphi_y). \qquad (1)$$

Here $E_{x,y}$ and $\varphi_{x,y}$ are the amplitudes and phases, respectively, of the two polarization components of the optical excitation pulse. The current flow is maximized for $\varphi_x = \varphi_y$ and $\varphi_x = \varphi_y + 180°$, i.e., for linearly polarized light with angles of +/- 45° to the $x$ direction of the sample. Switching between both excitation polarizations, which can be accomplished by changing the phase difference between two orthogonal polarization components, will result in a reversal of the current direction. For excitation of continuum transitions, the shift current's temporal behavior follows the temporal intensity profile of the excitation pulse[9].

In addition to the shift current contribution, diffusion and drift-based currents directed along the growth direction of the sample ($z$ axis) occur. On the one hand, differences in the electron and hole mobility of the photogenerated carriers affect the diffusion process and generate a net current flow in a direction normal to the surface, referred to as photo-Dember currents[11]. On the other hand, due to Fermi level pinning, the band structure near the surface is bent and an electrical field results. Optically generated carriers will be accelerated in this static surface field and lead to a current flow which is also directed normal to the surface[10]. For both current contributions, the rise time of the current is mainly given by the temporal length of the excitation pulse and the current direction does not depend on the polarization of the excitation pulse. Therefore, the emitted THz field does not change when the phase relation of the pump light is modified. Whether the photo-Dember effect or the surface-field effect dominates the current response along $z$, depends on the crystal properties and the excess energy of the excited carriers[12,13]. For the experimental conditions presented in this work, the surface-field effect has previously been shown to be dominant[12].

To further illustrate the steering effect, we schematically show the electric fields emitted from shift and surface-field currents in Fig. 1(a). For simplicity, refraction and diffraction effects are not included in this picture (although they will be treated in our theoretical model described below) and the point of view is restricted to the $y$-$z$ plane. In the left hand side of Fig. 1(a), $E_{Shift}$ corresponds to the THz signal emitted from the shift current, whereas $E_{SF}$ represents the THz signal emitted from the surface-field currents. Although the main emission direction of the dipole corresponding to the surface-field current is oriented in the $x$-$y$ plane, non-zero electrical field components have to be considered even for small deviations from the $z$ direction denoted by the angle β. Due to the rotational symmetry along the $z$ axis, the sign of the $y$ component of $E_{SF}$ is different on either side of the excitation spot in the $y$-$z$ plane. This leads to a constructive and destructive interference between $E_{SF}$ and $E_{Shift}$ at the left- and right-hand side, respectively, of the emission pattern. As a consequence the overall emission pattern will be tilted with respect to the $z$ direction. By changing the direction of the shift current, the interference effect and, consequently, the tilt angle can be reversed. This mechanism allows for coherently controlled steering of the direction of the overall THz emission.

For the experimental demonstration of this effect, an electro-optic sampling setup as shown in Fig. 1(b) has been used. The pump light was focused onto the (110)-oriented GaAs sample using a 150 mm lens. Depending on the distance between the lens and the sample this resulted in a $1/e^2$-diameter of the excitation spot ranging from approximately 125 μm to 400 μm. The average excitation power has been varied within a range between 100 mW to 280 mW. Although the pump-pulse polarization was adjusted by rotating a $λ/2$-plate, it should be emphasized that the polarization can also be rapidly adjusted by all-electronic (i.e., non-mechanical) means. The THz emission of the GaAs was detected using two 90 degrees off-axis parabolic mirrors (OPM) with 76.2 mm focal length and a ZnTe crystal (thickness 0.5 mm) for electro-optic detection. A THz polarizer placed between the sample and the first OPM ensured that only currents along $y$ contributed to the measured THz signal. In between the OPMs, a delay medium made of polypropylene was placed to analyze the emission direction as explained in the following.

We first demonstrate that a steering effect exists and quantify this effect later on. We placed the delay medium in between the OPMs such that it covered only the upper half of the OPM aperture, see inset of Fig. 2. By this, half of the THz signal was measured with a delay of about 2.3 ps. If the polarization direction of the excitation pulse was rotated from +45° to -45°, a change in the peak-to-peak THz signal occurred, see Fig. 2. The +45°-excitation polarization led to a smaller amplitude of the non-delayed THz signal as compared to the -45°-excitation polarization. This behavior was reversed for the delayed part of the THz signal, proving that it was not an improper alignment of the delay medium that caused this effect. For a further confirmation of this effect the GaAs was rotated around its $z$ axis by 90°. When exciting the $σ_{xyy}$-tensor element of the GaAs sample, which does not show a reversal of the current direction when switching the pump polarization as introduced above, no polarization-dependent change in the spatial field distribution has been measured. These observations prove that the THz power is not evenly distributed across the aperture in between the parabolic mirrors and this power distribution can be changed by switching the shift current direction. To investigate the spatial distribution in more detail, we fabricated a horizontal slit of 7 mm width into the delay medium (the same results were obtained with a metal slit) and measured the non-delayed peak-to-peak THz signal versus vertical slit position. Fig. 3 shows the resulting displacement of the THz signal for both excitation polarizations for 280 mW pump power and an excitation spot diameter of 125 µm. The measurement data has been corrected for the influence of the numerical aperture of the OPMs and demonstrates a clear and significant steering.

To exactly quantify the steering angles, a simple model is introduced. In this model, the temporal behavior of shift and surface-field currents is assumed to be comparable, both performing in-phase dipole-like oscillations. The corresponding THz field emitted from each current contribution is modeled using standard dipole radiation patterns in the far-field approximation[14].

$$\boldsymbol{E_{Dipole}} = \frac{1}{4\pi\varepsilon_0}\frac{k^2 \exp(ikr)}{r}(\hat{\boldsymbol{n}} \times \boldsymbol{p}_{Shift/SF}) \times \hat{\boldsymbol{n}}. \qquad (2)$$

Here, $\varepsilon_0$ denotes the electric constant, $k$ is the wave number, $r$ is the distance from the dipole to the observation point, $\hat{\boldsymbol{n}}$ denotes the unit vector in direction to the observation point and $\boldsymbol{p}_{Shift/SF}$ represents the dipole moment of each current contribution, with $\boldsymbol{p}_{Shift}$ and $\boldsymbol{p}_{SF}$ being oriented along the $y$ and $z$ direction, respectively. Evaluating Eq. (2) for both dipoles and superimposing their electrical field vectors, the overall emission pattern is obtained. We also consider diffraction effects resulting from the size of the excitation spot diameter. Since the dimension of the spot is on the order of the center THz wavelength, the focal spot acts as a circular aperture and in the far field the well known Fraunhofer diffraction pattern results[15]. In addition to diffraction, polarization-depending refraction occurring at the GaAs/air interface (refraction index $n_{GaAs} = 3.6$) is included. This dipole model can be used to approximate the measured data by optimization of only two fit parameters: (i) the ratio between the strength of surface-field and shift currents, denoted by the ratio of the absolute values of the dipole moments, $a_1 = p_{SF}/p_{Shift}$ and (ii) the influence of diffraction depending on the ratio between the center THz wavelength $\lambda$ and excitation spot size $d$, $a_2 = d/\lambda$. By optimizing the values for $a_1$ and $a_2$ using a least-squares algorithm, the model shows an excellent agreement to the experiment, see solid lines in Fig. 3. From the fit to the experiment we extract a displacement of the maximum THz fields shown in Fig. 3 of ~10 mm. This is equal to an angular change of the emission direction of the THz signal of 8 degrees. To verify this result, the measurements have been repeated using (113)-oriented GaAs. Here, similar steering effects are obtained along both in-plane directions [1$\bar{1}$0] and [33$\bar{2}$]. This is expected from symmetry considerations since the shift current directions along both crystal axes can be coherently controlled by adjusting the polarization of the pump light.

We have also studied the dependence of the steering effect on pump power and excitation spot size. As depicted in Fig. 4, these dependencies can be used to adjust the steering angle in a wide range. Fig. 4(a) shows the measured steering angle versus pump power for a fixed excitation spot size of 180 µm. Varying the power from 100 mW to 260 mW (variation of excitation intensity from 50 MW/cm$^2$ to 135 MW/cm$^2$) results in an increase of the steering angle from 2 degrees to 6 degrees. Fig. 4(c) shows the corresponding fit parameters $a_1$ and $a_2$ that were obtained for each measurement. From this graph the variation of the steering effect can clearly be attributed to the dipole moment ratio $a_1$, which increases with increasing pump power. We attribute this effect to different saturation behaviors of shift and surface-field currents. The shift current shows strong saturation effects at high excitation intensities[16,17], whereas the saturation for surface-field currents is expected to be much smaller[18,19]. In contrast to $a_1$, the fit parameter $a_2$ remains constant for different pump powers showing that no change in the diffraction properties takes place.

When varying the spot size for a fixed pump power of 280 mW, a more complex behavior is obtained, refer to Figs. 4(b) and (d). The decrease of the excitation spot diameter from 375 μm to 125 μm (variation of excitation intensity from 30 to 300 MW/cm$^2$) results in an increase of the steering angle from below 3 degrees to 8 degrees. In this case both fit parameters change, see Fig. 4(d). Increasing the excitation spot diameter for constant pump power goes along with a decreasing power density and, consequently, as described above $a_1$ decreases. Moreover, due to the change of the spot size, diffraction effects vary as well. This leads to a decrease of the fit parameter $a_2 = d/\lambda$ with decreasing spot size. In fact we obtain a sublinear dependence of $a_2$ on $d$ (decreasing $d$ by a factor of 3 results in a decrease of $a_2$ by a factor of about 1.3), which we attribute to wavelength-dependent diffraction, i.e., the center wavelength $\lambda$ depends strongly on $d$, too[20]. Consequently an increase of the THz center frequency is expected for the given reduction of $d$. We confirm this dependence by measurements of THz spectra, which qualitatively show a stronger suppression of lower frequency components for smaller spot sizes (not shown here).

Finally, another relevant factor influencing the steering angle has to be pointed out. Our simulations show that a reduction of the refraction-index mismatch at the semiconductor-air interface, i.e., a reduction of the total internal reflection at the backside of the semiconductor crystal, might significantly enhance the steering effect. In principle, this can be accomplished by attaching a hemispherical Si lens to the sample. However, care has to be taken in determining appropriate dimensions of the Si lens such that the lens does not focus the emitted THz radiation, which would be counterproductive and reduce the steering effect. Such a study is out of the scope of this work.

In summary, we have demonstrated THz beam steering by optical coherent control. By reversing the direction of shift currents and superimposing the THz radiation emitted from these currents with THz radiation emitted from surface-field currents, steering angles of up to 8 degrees have been realized. As shown in a simple dipole-based model including refraction and diffraction effects, the size of the steering angle mainly depends on the relative strength of both current contributions and on diffraction effects due to the excitation spot size. Due to this dependence the THz steering effect may even be utilized to further investigate carrier transport dynamics of shift and surface-field currents.

**Figure 1**

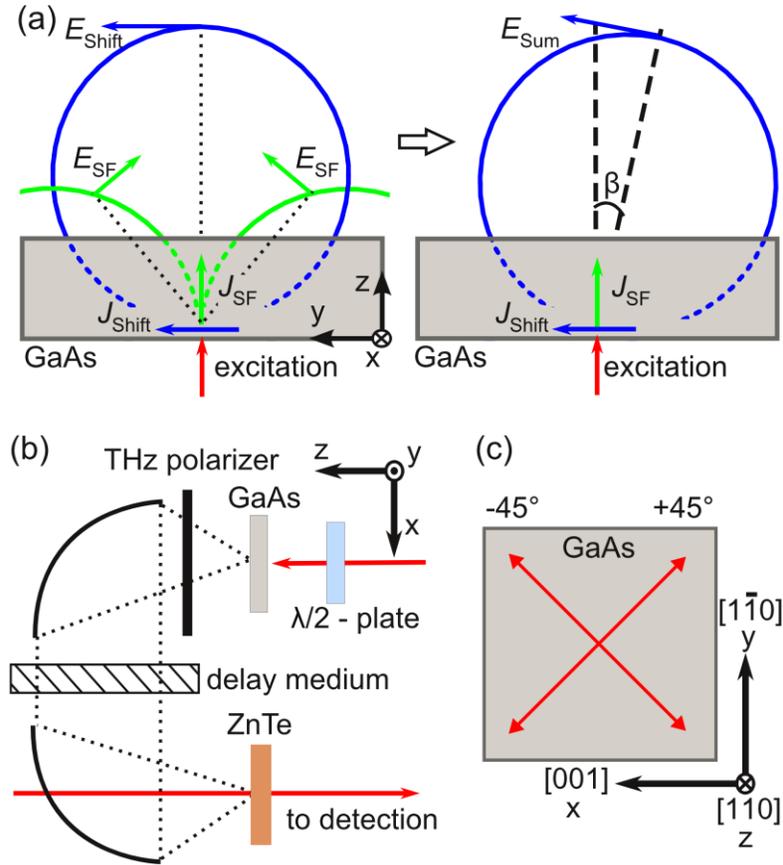

Fig. 1: (a) Dipole radiation resulting from shift and surface-field currents, $E_{Shift}$ and $E_{SF}$, respectively. The superposition of these fields results in a tilted direction of the overall THz emission. (b) Electro-optic detection setup (top view). (c) Polarization directions of the

**Figure 2**

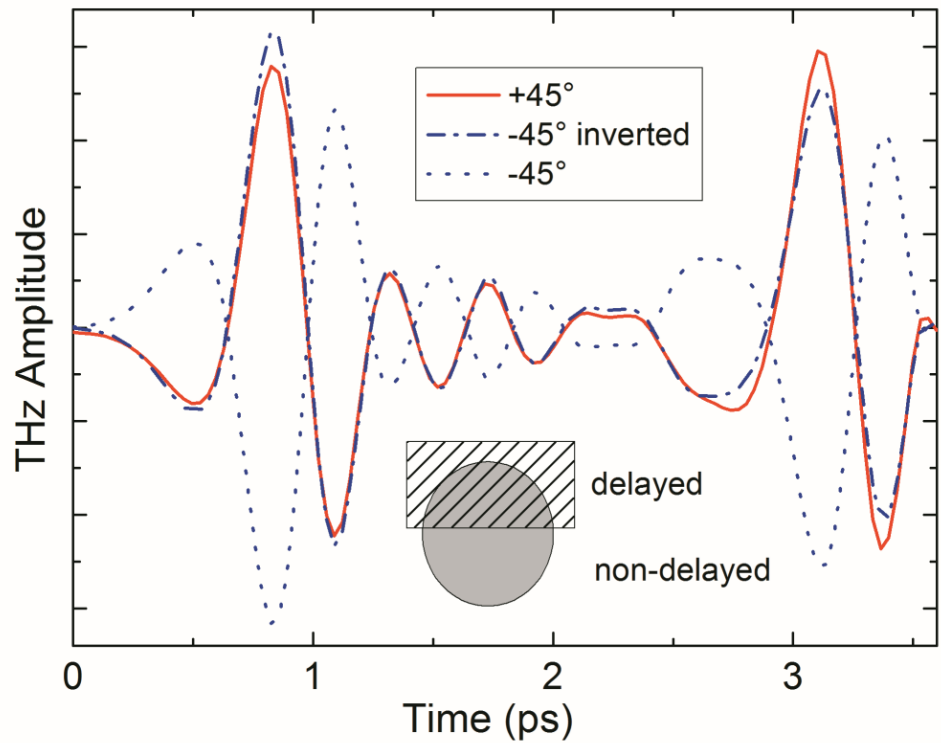

Fig. 2: THz traces emitted from the (110)-oriented GaAs sample for excitation polarization having an angle of ±45° (solid / dotted line) with respect to the *x* axis of the crystal and a delay medium in between the parabolic mirrors. To allow for a better comparison of amplitudes the dotted line has been inverted (dash-dotted line).

**Figure 3**

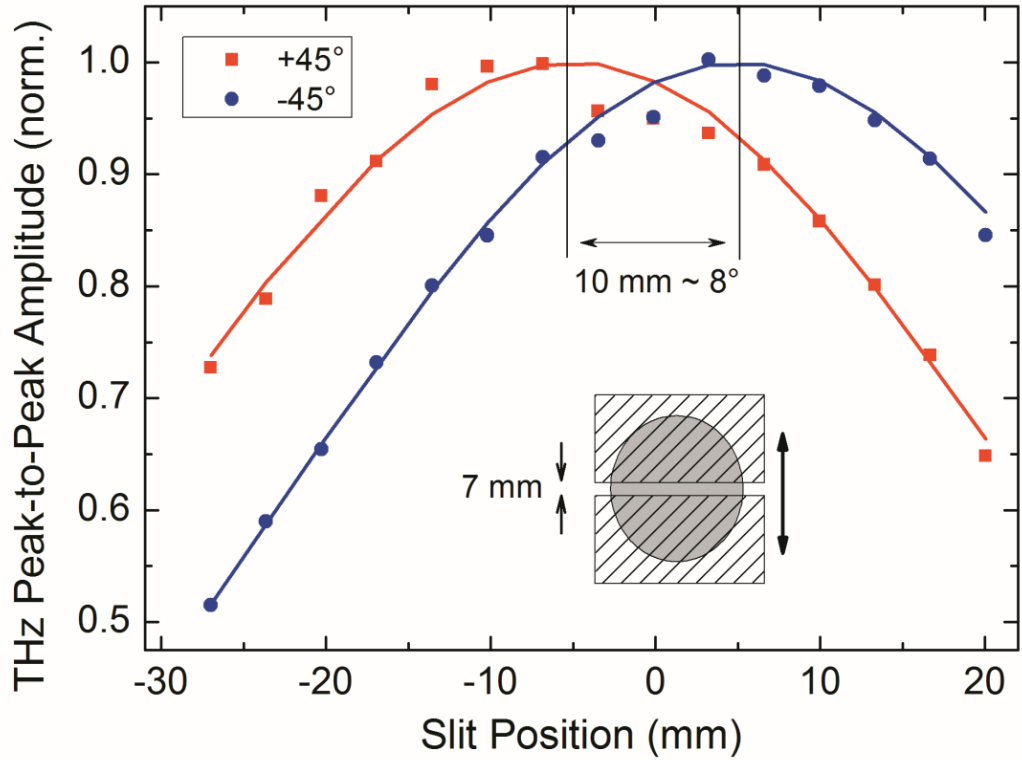

Fig. 3: Peak-to-peak THz signal transmitted through a 7 mm wide slit versus slit position. By changing the excitation polarization from -45° to +45° a displacement of the THz beam pattern of ~10 mm is obtained. This corresponds to a steering angle of 8 degrees.

**Figure 4**

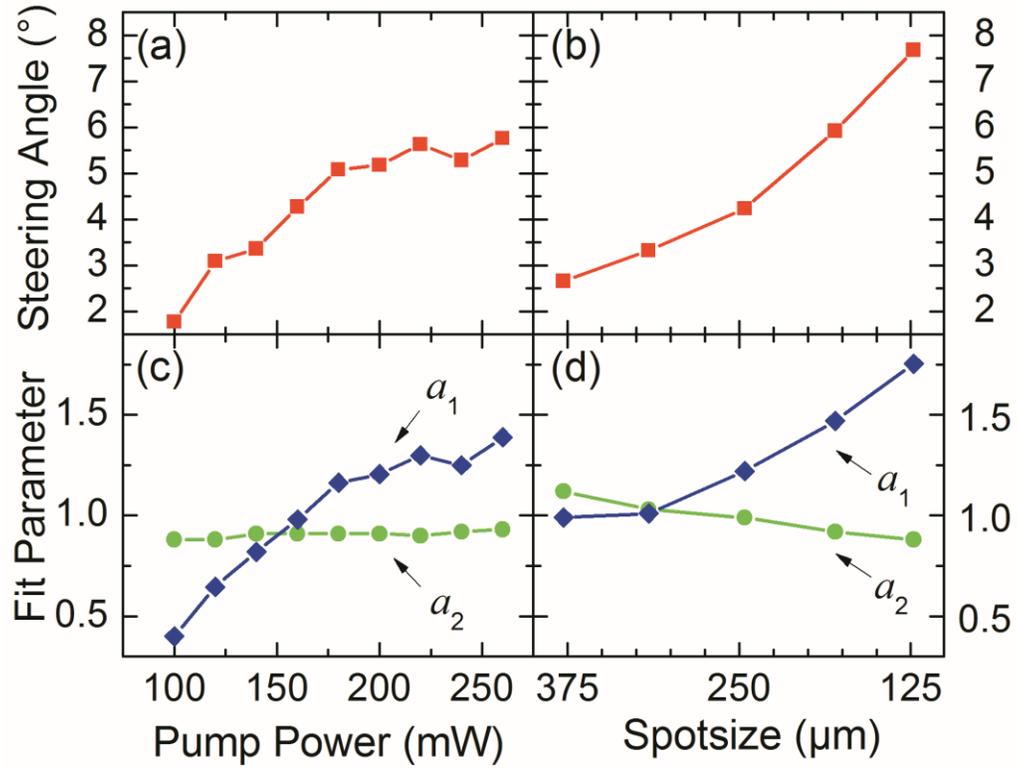

Fig. 4: (a) Steering angle versus pump power for a fixed excitation spot size of 180 μm. (b) Steering angle versus $1/e^2$ excitation-spot diameter for a fixed excitation power of 280 mW. (c),(d) Parameters $a_1$ (diamond) and $a_2$ (dot) obtained from a fit of the analytical model to the measurements shown in (a), (b).